\documentclass[onecolumn,preprintnumbers,amsmath,amssymb,prl,superscriptaddress]{revtex4}

\usepackage[paperwidth=210mm, paperheight=297mm,centering, hmargin=3.2cm, vmargin=2.5cm]{geometry}
\usepackage{graphicx}
\usepackage{epstopdf}
\usepackage{dcolumn}
\usepackage{bm}
\usepackage{amsmath}
\usepackage{amssymb}
\usepackage{mathrsfs}
\usepackage{amsfonts}
\usepackage{times}
\usepackage{subfigure}
\usepackage{amstext}
\usepackage[colorlinks=true, citecolor=blue]{hyperref}
\usepackage[dvipsnames]{xcolor}
\usepackage{ulem}

\newcommand{\ket}[1]{|#1\rangle}

\begin{document}

\title{Observing the Spectral Collapse of Two-Photon Interaction Models\footnote{Presented at the 11th Italian Quantum Information Science Conference (IQIS2018), Catania, Italy, 17–20 September 2018.} }
\author{Simone Felicetti}
\affiliation{Laboratoire Mat\'eriaux et Ph\'enom\`enes Quantiques,
Universit\'e de Paris, CNRS, 75013 Paris, France}
\author{Alexandre Le Boit\'e}
\affiliation{Laboratoire Mat\'eriaux et Ph\'enom\`enes Quantiques,
Universit\'e de Paris, CNRS, 75013 Paris, France}

\begin{abstract}Until very recently, two-photon interaction processes have been considered only as arising from second- or higher-order effects in driven systems, and so limited to extremely small coupling strengths. However, a variety of novel physical phenomena emerges in the strong and ultrastrong coupling regimes. 
Strikingly, for a critical value of the coupling strength the discrete spectrum collapses into a continuous band. In this extended abstract, we discuss recent proposals to implement genuine two-photon interactions in an undriven solid-state system, in the framework of circuit QED. In particular, we review counterintuitive spectral features of two-photon interaction models and we show how  the onset of the spectral collapse can be observed in feasible scattering experiments.
\end{abstract}
    
\maketitle 
 


\vspace{-0.5cm}

\hrulefill

\subsection{Introduction}

Quantum technologies provide a perfect tool to implement minimal models of light-matter interaction in a controllable way. In particular, in the framework of cavity and circuit quantum electrodynamics (QED), the spatial confinement of the electromagnetic field enhances the light-matter coupling strength making it possible to explore extreme regimes of interaction. 
In the strong-coupling (SC) regime, where the coupling strength overcomes the decay and decoherence rates, quantum effects such as Rabi oscillations become observable.
When the light-matter interaction strength is increased further up to values comparable with the bare system frequencies, the ultrastrong-coupling (USC) regime is reached. In this regime, the rotating-wave approximation (RWA) ceases to be applicable and a much more complex physics emerges. The USC regime has been experimentally achieved in different quantum platforms~\cite{Diaz2018}. Among those, the circuit QED technology stands out for its flexibility in the implementation of artificial two-level atoms, namely superconducting qubits, that can interact with bosonic modes supported by microwave resonators. Since the first experimental demonstrations, the interest in the USC regime has been steadily growing, concerning both fundamental properties and quantum-information applications. Beyond that, the achievement of the USC regime has fostered the interest in different generalization of minimal quantum optical models. 

In this context, two-photon-interaction (TPI) models are particularly interesting due to their highly counter-intuitive spectral features. In particular, in the ultrastrong coupling regime a spectral collapse takes place:
for a specific value of the coupling strength the discrete spectrum collapses into a continuous energy band. 
Recently, TPI models have been at the center of an intense theoretical research aimed at characterizing their spectral 
properties~\cite{Travenec2012,Albert2011,Duan2016,Cui2017,Lupo2017,Ying2018,Duan2018,Cong2018}, particularly in proximity of the spectral collapse.
In the many-body limit, it has been shown that TPIs lead to a rich phase diagram characterized by the interplay between the spectral collapse and the superradiant phase transition~\cite{Garbe2017,Chen2018}.
The debate is still open on whether such spectral collapse is observable experimentally or simply an artefact of an incomplete theoretical description.

A key challenge to experimental verifications is that the implementation of TPI requires an interaction more complex than the usual dipolar coupling. So far, effective TPI have been realized only using second- or higher-order effects of the dipolar interaction in driven systems~\cite{DiPiazza2012} and therefore limited to extremely small coupling strengths. To overcome this problem, different quantum-simulation schemes have been proposed to effectively reproduce the physics of ultrastrong TPI in atomic systems~\cite{Felicetti2015,Puebla2017,Schneeweiss2017,Cheng2018}. Effective implementations of multiphoton interaction models have also been considered as a tool for quantum simulations~\cite{Pedernales2018} and quantum-information processing~\cite{Rossatto2018}. 

Going beyond quantum simulation proposals, recent theoretical analyses~\cite{Ref1,Ref2} show that ultrastrong TPI could be genuinely implemented in an undriven solid-state device. In particular, it was shown that in the context of circuit QED it is possible to engineer an intrinsically nondipolar coupling between a flux qubit and a microwave resonator. In this way, dipolar and two-photon interaction terms can be selectively activated~\cite{Ref1} and, in the case of galvanic coupling, the interaction strength can be pushed into the USC regime~\cite{Ref2}.
In the following,  we will briefly introduce the spectral properties of TPI models and we will show how they affect the optical response of the system. In particular, we will describe how a feasible scattering experiment would allow to witness the transition from the strong to the ultrastrong coupling regime of TPI, and to observe a direct signature of the onset of the spectral~collapse.

\subsection{The Spectral Collapse}
\begin{figure}[]
\includegraphics[width=\textwidth]{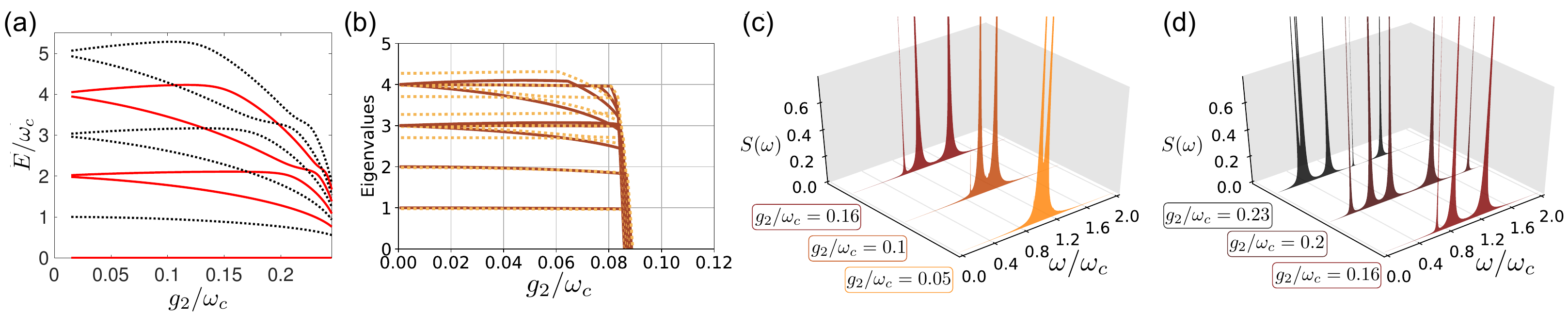}
\caption{ \label{Fig1}
 (\textbf{a},\textbf{b})
Energy spectrum for the two-photon Dicke model~\eqref{twoPhDicke} as a function of the coupling strength, for atoms resonant with the two-photon transition $\omega_q = 2\omega_c$.
Panel (\textbf{a}) shows~\cite{Ref2} the spectrum  for $N=1$; different line styles denote the parity of the number of photons (positive for continuous red lines, negative for dotted black lines),  which is a symmetry of ${\hat H}_{2ph \rm}$.
Panel (\textbf{b}) shows~\cite{Ref1} the spectrum  for $N=3$, the solid red line corresponds to the two-photon Dicke  ${\hat H}_{2ph \rm}$, while the yellow dashed line corresponds to the 
 model obtained including inter-spin couplings $J \sum_{i=1}^{N-1} \sigma_x^i\sigma_x^{i+1}$ of strength $J = 0.2\omega_c$.
  (\textbf{c},\textbf{d}) Fluorescence spectrum~\cite{Ref2} in arbitrary units of the output signal under coherent drive, for different values of the coupling strength. In (\textbf{c}) is shown the transition from the strong to the USC regime, while in  (\textbf{d}) the fluorescence spectrum bears a direct signature of the onset of the spectral collapse.}
\end{figure}

Consider a system composed of an ensemble of $N$ two-level atoms, or qubits, interacting with a single cavity mode via TPIs. This model is known as the two-photon Dicke model and its Hamiltonian can be written as
\begin{equation} 
\label{twoPhDicke}
{\hat H}_{2ph \rm} = \omega_c \hat a^\dagger \hat a + \frac{\omega_q}{2}\sum_{i=1}^N \hat  \sigma^{(i)}_z + g_2 \sum_{i=1}^N \hat  \sigma^{(i)}_x \left( \hat a^\dagger + \hat a  \right)^2,
\end{equation}
where we defined the cavity $\omega_c$ and qubit $\omega_q$ frequency, and the two-photon coupling strength $g_2$. Notice that in the literature the TPI term is often written as 
$\hat  \sigma^{(i)}_x \left( \hat a^\dagger a^\dagger + \hat a \hat a  \right)$, whereas we take here the interaction Hamiltonian to be in the full-quadratic form. The latter form of the interaction arises when the qubit is coupled to the square of the electric (or magnetic) field of the cavity mode and it is therefore a natural generalization of the standard dipolar coupling~\cite{Ref1}.
The system spectrum is shown in Figure~\ref{Fig1}a,b for $N=1$ and $N=3$, as a function of the coupling strength $g_2$.  The spectral collapse takes place for the critical value of the two-photon coupling strength $ g_{\text{col}}= \frac{\omega_c}{4N} $, which depends only on the cavity frequency and on the number of atoms. At the collapse point, the spectrum is composed of a finite number of discrete levels and of a continuous energy band~\cite{Travenec2012,Albert2011,Duan2016,Cui2017,Lupo2017,Ying2018,Duan2018}. For larger values of the coupling strength the spectrum is unbounded from below and thus the model is 
no longer well defined.  In Figure~\ref{Fig1}b we plot the spectrum also in the case in which direct inter-atomic interactions $J \sum_{i=1}^{N-1} \sigma_x^i\sigma_x^{i+1}$  have been included (dashed yellow line). This shows that qubit-qubit interactions do not affect the onset of the spectral collapse, in contrast with the superradiant phase transition of the standard Dicke model, where inter-atomic couplings inevitably inhibit the transition~\cite{Jaako16}.

\subsection{Fluorescence Spectrum}
Let us now show in detail how such spectral properties could be observed.
We consider the $N=1$ case and we assume that both the cavity and the  atom can be coupled to open transmission lines to implement an optical drive and to read the output signal. The numerical results~\cite{Ref2} shown in the following have been obtained using an input-output theory which is valid for all values of the coupling strength~\cite{Ridolfo2012,LeBoite2017}. We denote with $\ket{\Psi_n^\pm}$ the $n$-th excited state in the $\pm1$ parity subspace. We consider the input-output scheme where the atom is driven with a coherent drive resonant with the transition $\ket{\Psi_0^+} \to \ket{\Psi_2^+}$, i.e., from the ground state to the second excited state in the even parity subspace [red continuous lines in Figure~\ref{Fig1}(a)]. Finally, we assume that the output signal emitted through the cavity is measured. We focus on the fluorescence spectrum, namely, the frequency components of the emitted field.
The fluorescence spectrum can be calculated as the average over $t$ of $S(\omega, t) =  \int_{-\infty}^{+\infty} d \tau e^{i\omega \tau}g(t,t+\tau)$,  where $g(t,t+\tau)$ is the Fourier transform of a two-time correlation function of the output field (see Ref.~\cite{Ref2} for details).

The numerically-computed fluorescence spectrum is shown in Figure~\ref{Fig1}c,d, for different values of the coupling strength. In Figure~\ref{Fig1}c, we show the transition from the SC to the USC regime. 
 For~$g_2/\omega_c = 0.005$ the spectrum has only two peaks corresponding to the transitions $\ket{\Psi_2^+} \to \ket{\Psi_0^-}$ and $\ket{\Psi_0^-} \to \ket{\Psi_0^+}$, which is what we expect in the regime where the selection rules arising under the RWA are valid.  On the other hand, as the coupling is increased ($g_2/\omega_c = 0.1$ and $g_2/\omega_c = 0.165$) a third resonance at the frequency of the transition $\ket{\Psi_1^+} \to \ket{\Psi_0^-}$ appears. Such line can be observed only if the transition $\ket{\Psi_2^+} \to \ket{\Psi_1^+}$ is no longer forbidden by selection rules, and so the appearance of this additional peak is a direct witness of the breakdown of the rotating-wave approximation. 
 
Finally, in Figure~\ref{Fig1}d, we show the behavior of the fluorescence spectrum as the value of the two-photon interaction strength is increased further. When $g_2$ approaches the collapse point, two dramatic changes  can be observed. 
First, multiple additional peaks emerge when $g_2/  \omega_c \gtrsim 0.17$, due~to additional decay channels that become available after the level crossing observable in the energy spectrum for this coupling value [see Figure~\ref{Fig1}(a)]. Second, and more importantly, a global red-shift is observable in the whole fluorescence spectrum. Namely,  peaks in $S(\omega)$ get closer together for increasing values of the coupling. To conclude, the multi-peak pattern in a narrow frequency window of the fluorescence spectrum represents a clear signature of the onset of the spectral collapse.

\subsection{Conclusions}
To conclude, the implementation of genuine two-photon interactions in a solid-state system would make it possible to directly investigate the counter-intuitive spectral properties of ultrastrong TPI models in feasible scattering experiments. In particular, via fluorescence spectroscopy it could be demonstrated that the spectral collapse is not only an interesting mathematical property, but a genuine physical phenomenon.

\vspace{6pt} 
\acknowledgments{S.F. acknowledges support from the French Agence Nationale de la Recherche (SemiQuantRoom, Project No. ANR14-CE26-0029) and from the PRESTIGE program, under the Marie Curie Actions---COFUND of the FP7.}








\end{document}